\documentclass[conference]{IEEEtran}
\usepackage[left=0.64in, right=0.64in, top=0.73in, bottom=0.5in, includefoot]{geometry}
\usepackage{graphicx}
\usepackage{subcaption}
\usepackage{url}
\usepackage{array}
\usepackage[utf8]{inputenc} 
\usepackage[T1]{fontenc}
\usepackage{amsmath}
\usepackage{mathtools}
\usepackage{blkarray}
\usepackage{amsfonts}
\usepackage{amssymb}
\usepackage{amsthm}
\usepackage{enumerate}
\usepackage{cite}
\usepackage{lipsum}
\usepackage{mathtools}
\usepackage{cuted}
\usepackage{calc}
\usepackage{color}
\usepackage{epsfig}
\usepackage{setspace}
\usepackage{pstricks}
\usepackage{cancel}
\usepackage{multirow}
\usepackage{mathrsfs}
\usepackage{algorithm}
\usepackage{algpseudocode}
\usepackage{multirow}
\usepackage{diagbox}
\usepackage{multicol}
\usepackage{booktabs,makecell}
\usepackage{mathrsfs}

\usepackage{enumitem}

\usepackage{enumitem}
\newtheorem{defn}{Definition}
\newtheorem{thm}{{\cal T}heorem}

\newtheorem{constr}{Construction}

\newtheorem{example}{Example}

\newcommand{\R}{\mathcal{R}}

\newcommand{\M}{\mathcal{M}}

\newcommand{\U}{\mathcal{U}}

\newcommand{\J}{\mathcal{J}}

\input{epsf.sty}

\begin{document}
	\title{Privacy-Preserving Coding Schemes for Multi-Access Distributed Computing Models}
	\author{
		\IEEEauthorblockN{Shanuja Sasi\\}
		\IEEEauthorblockA{Department of Electrical Engineering, Indian Institute of Technology Kanpur, India \\
			E-mail: shanujas@iitk.ac.in}
	}
	\maketitle
	\begin{abstract}

Distributed computing frameworks such as MapReduce have become essential for large-scale data processing by decomposing tasks across multiple nodes. The multi-access distributed computing (MADC) model further advances this paradigm by decoupling mapper and reducer roles: dedicated mapper nodes store data and compute intermediate values, while reducer nodes are connected to multiple mappers and aggregate results to compute final outputs. This separation reduces communication bottlenecks without requiring file replication.  In this paper, we introduce privacy constraints into MADC and develop private coded schemes for two specific connectivity models. We construct new families of extended placement delivery arrays and derive corresponding coding schemes that guarantee privacy of each reducer’s assigned function.
	
	\end{abstract}
	
	\section{Introduction}
Distributed computing (DC) methods are fundamental for scaling modern machine learning and large-scale data processing tasks by decomposing complex computations into parallelizable subtasks across multiple nodes. Frameworks like MapReduce \cite{Mapreduce} exemplify this paradigm, enabling efficient processing of massive datasets by dividing work into \textit{Map}, \textit{Shuffle}, and \textit{Reduce} phases. Initially, a \textit{Map phase} is responsible for tasks such as feature extraction and transformation. During this phase, each node processes a subset of the data, extracting relevant features and producing intermediate outputs. Following this, a \textit{Shuffle phase} organizes these intermediate results based on specific criteria, such as feature indices or class labels. This step ensures that related data is efficiently routed to nodes tasked with subsequent operations. Finally, a \textit{Reduce phase} consolidates the processed outputs, combining intermediate results to compute final output functions, such as gradients or model predictions.  A critical challenge in these systems, however, is maintaining data and process {privacy}, especially in sensitive applications like federated learning or multi-tenant cloud environments. 
 
 Coding-theoretic techniques have been extensively applied in  DC for a wide range of applications, including caching \cite{MAN,DP,DP2}, coded matrix multiplication \cite{MCJ}, gradient computations \cite{TLDK,SAR}, and coded MapReduce \cite{LMA,YYW,SFZ,WCJ,WCJnew,BP,SGR,sasi  topology,sasi secure t desi,sasi rate,sasi secure con,sasi secure journal}. Some of these works leverage {\it placement delivery arrays (PDAs)} \cite{JQ,YTC}. 
In \cite{LMA}, coding-theoretic techniques have been successfully integrated into DC to optimize the trade-off between computation and communication loads, notably through {\it coded distributed computing (CDC)}. Recent work has introduced {\it private CDC}, which incorporates privacy constraints to conceal the specific output function index assigned to each computing node in DC framework \cite{sasi private isit,sasi private tcom}. 

A new computing framework called \textit{multi‑access distributed computing (MADC)} model was introduced in \cite{BP}. In contrast to the classical model of \cite{LMA} where every node acts both as a mapper and a reducer, the MADC model decouples these roles: mapper nodes store input data and generate intermediate values, while reducer nodes collect intermediate values from the mapper nodes they are connected to, exchange information among themselves, and finally compute the output functions.
The MADC model has been further investigated in \cite{SGR,sasi  topology,sasi secure t desi}, where several network topologies were studied and corresponding coded schemes were developed. A key observation from \cite{SGR} is that, by exploiting the specific connectivity between mappers and reducers, the communication load can be reduced even without file replication, which is an advantage not available in the conventional MapReduce framework, where  CDC fundamentally relies on data replication to lower communication overhead. Such MADC architectures are especially well‑suited for edge‑computing and data‑center environments, where multi‑access links help alleviate communication bottlenecks.
Existing private CDC schemes \cite{sasi private isit,sasi private tcom}, although effective, assume a conventional DC framework. A notable gap therefore remains: \textit{how to preserve task‑assignment privacy in a MADC setting}, where each reducer is linked to several mappers and the associations among mappers, reducers, and computed functions could leak sensitive operational patterns. 

\noindent {\it Our Contributions}:
 In this paper, we incorporate  a privacy constraint into MADC to ensure that the task (i.e., output function index) assigned to each reducer node remains hidden from others. This guarantees that no node can infer another's specific role or data characteristics, which preserves privacy. We refer MADC models with privacy constraints as \textit{private coded MADC models}.  We focus on two specific connectivity patterns within this framework and construct a new class of PDAs to provide coding schemes that meet these privacy requirements. Under the imposed privacy constraints, we characterize achievable computation and communication loads. Importantly, we aim to minimize both loads without resorting to file replication, i.e., we achieve a computation load of 
$1$ while maintaining privacy.

\noindent {\it Organization of this paper: }	The MADC framework is formally defined in Section \ref{problem definition}. Subsequently, Section \ref{dlmodel} introduces the {\it private coded MADC model}, where we present the two specific models examined in this work: the \textit{$\alpha$-connect Partial Private Mapper Assignment-MADC Model} and the \textit{$\alpha$-cyclic Partial Private Mapper Assignment-MADC Model}. Both models are defined within this section. In Sections \ref{main} and \ref{main2}, we construct new extended PDAs and provide private coded schemes for these models, presented in Theorems 1 and 2, respectively.

\noindent {\it Notations:}
 $[a,b]$ denotes the set $\{a,a+1,a+2,\dots,b\}$ of consecutive integers from $a$ to $b$, $[a,b)$ denotes the set $\{a,a+1,a+2,\dots,b-1\}$ and $\oplus$ stands for the bitwise XOR operation. The $i$-th entry of a vector $A$ is written as $A[i]$. Finally, for integers $a$ and $b$, we define the bracket operation $[a]_b$ by $[a]_b = a$ if $b \ge a$, and $[a]_b = a-b$ otherwise. The set $\{[a]_b, [a+1]_b, \dots, [a+c]_b\}$ is represented by $[a, c]_b$.
	\section{Problem Definition}
	\label{problem definition}
	
	\begin{figure}
		\centering
		\includegraphics[scale=0.5]{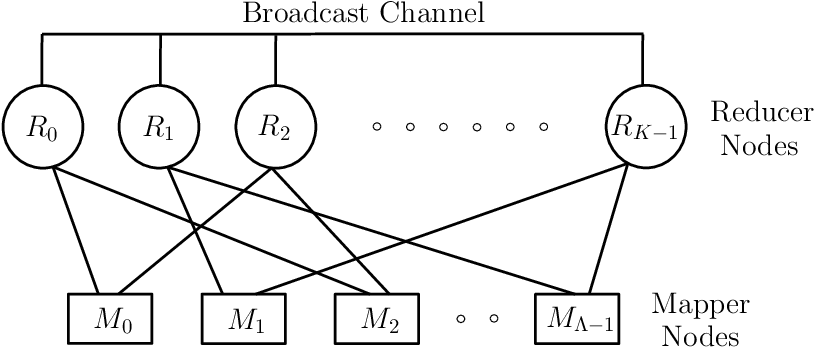}
		\caption{An MADC model with $\Lambda$ mapper nodes  and $K$ reducer nodes.}
		\label{fig: madc model}
	\end{figure}
	
	We consider a MADC framework consisting of \(\Lambda\) mapper nodes \(\{M_{\lambda}:\lambda \in [\Lambda]\}\) and \(K\) reducer nodes \(\{R_k: k \in [K]\}\) (as shown in Fig. \ref{fig: madc model}).  
	Each reducer node \(R_k\) is connected to a subset of mapper nodes and is assigned to compute exactly one of \(Q\) output functions \(\{\phi_q: q \in [Q]\}\).  
	Every output function \(\phi_q\) depends on \(N\) input files \(\{W_n : n \in [N]\}\).
	
	An input file \(W_n \in \mathbb{F}_{2^d}\) (for \(n \in [N]\)) contains \(d\) bits.  
	The output function \(\phi_q\) maps all \(N\) files to a bit\-stream of length \(b\); i.e., \(\phi_q : \mathbb{F}_{2^d}^N \rightarrow \mathbb{F}_{2^b}\).  
	For each \(n \in [N]\), a map function \(g_{q,n} : \mathbb{F}_{2^d} \rightarrow \mathbb{F}_{2^{\beta}}\) converts file \(W_n\) into an intermediate value (IV) \(v_{q,n} = g_{q,n}(W_n) \in \mathbb{F}_{2^{\beta}}\) of \(\beta\) bits.  
	A reduce function \(h_q : \mathbb{F}_{2^{\beta}}^N \rightarrow \mathbb{F}_{2^b}\) then aggregates all IVs to produce the final output \(h_q(v_{q,1},\dots,v_{q,N}) \in \mathbb{F}_{2^b}\) of \(b\) bits.  
	Hence, each output function can be written as $
	\phi_q(W_1,\dots,W_N) = h_q(v_{q,1},\dots,v_{q,N}),$ for $ q \in [Q].$
	 The  computation follows three sequential phases:
	
	\begin{enumerate}
		\item \textbf{Map Phase:}  
		The \(N\) input files are divided into \( \Lambda\) disjoint batches \(\mathcal{B} = \{B_1, B_2, \dots, B_{\Lambda}\}\), each containing \(\eta_1 = N/\Lambda\) files, so that \(\bigcup_{m=1}^{\Lambda} B_m = \{W_1, \dots, W_N\}\).  
		Each mapper node \(M_\lambda\) (with \(\lambda \in [\Lambda]\)) stores exactly one batch \(B_\lambda\).  
		It computes all IVs for that batch, i.e., $
		\{v_{q,n} = g_{q,n}(W_n) : q \in [Q],\; W_n \in B_\lambda\},$
		where each \(v_{q,n}\) is a bitstream of length \(\beta\).
		
		\item \textbf{Shuffle Phase:}  
		Each reducer node \(R_k\) (\(k \in [K]\)) is assigned a single output function \(\phi_{d_k}\) to compute.  
		Based on its connectivity, \(R_k\) can retrieve IVs directly from the mapper nodes it is connected to.  
		To obtain the remaining IVs required for its function, the reducer nodes cooperate: each reducer \(R_k\) constructs a coded message \(\mathbf{X}_k \in \mathbb{F}_{2^{l_k}}\) from the IVs it possesses and broadcasts it to all other reducers over an error‑free broadcast channel.
		
		\item \textbf{Reduce Phase:}  
		Using the received broadcast sequences \(\{\mathbf{X}_j\}_{j \in [K] \setminus k}\), each reducer \(R_k\) decodes all IVs needed to compute \(\phi_{d_k}\).  
		It then evaluates its assigned output function.
	\end{enumerate}
	
	The performance of a MADC scheme is quantified by two metrics that we aim to optimize.
	
	\begin{defn}[Computation Load]
		The computation load \(r\) is the total number of files mapped across the mapper nodes, normalized by the total number of files \(N\).  
		This represents the average replication factor of each file.
	\end{defn}
	
	\begin{defn}[Communication Load]
		The communication load \(L\) is the total number of bits transmitted by all reducers during the Shuffle phase, normalized by the aggregate bit‑length of all IVs.  
		It measures the communication overhead required for reducers to exchange the missing IVs.
	\end{defn}

\subsection{Placement Delivery Array Background}
Yan et al. \cite{YCTCPDA} introduced the concept of PDAs as a framework for designing coded caching schemes with reduced sub-packetization requirements. Following this work, numerous coding schemes based on PDAs have been developed.
\begin{defn}  ({\bf Placement Delivery Array}\cite{YCTCPDA}):
	\label{def pda}
	For positive integers $K, F, Z,$ and $S,$ an $F \times K$ array $P = [p_{f,k}]$ with $ f \in [F],$ and $ k \in [K]$ composed of a specific symbol $*$ and $S$ positive integers $[S],$ is called a $(K, F, Z, S)$ placement delivery array (PDA) if it satisfies the following conditions:
	\begin{itemize}
		\item {\it A1:} The symbol $*$ appears $Z$ times in each column;
		\item {\it A2:} Each integer occurs at least once in the array;
		\item {\it A3:} For any two distinct entries $p_{f_1,k_1}$ and $p_{f_2,k_2}, s=p_{f_1,k_1} = p_{f_2,k_2} $ is an integer only if
		\begin{enumerate}
			\item $f_1$ $\neq f_2$ and $ k_1$ $\neq k_2,$ i.e., they lie in distinct rows and distinct columns; and
			\item $p_{f_1,k_2} = p_{f_2,k_1} = *,$ i.e., the corresponding $2 \times 2$ sub-array formed by rows $f_1, f_2$ and columns $k_1, k_2$ must be either of the following forms
			$ \begin{pmatrix}
			s & *\\
			* & s
			\end{pmatrix} $ or 
			$\begin{pmatrix}
			*& s\\
			s & *
			\end{pmatrix}.$\qed
		\end{enumerate} 
	\end{itemize}
	\label{def:PDA}
\end{defn}
\begin{defn} ({\bf $g-$regular PDA}\cite{YCTCPDA}):
	An array $P$ is said to be a $g-(K, F, Z, S)$ PDA if it satisfies A1, A3, and the following condition
	\begin{itemize}
		\item {\it $A2'$:}  Each integer appears $g$ times in $P$, where $g$ is a constant.
	\end{itemize}
	\label{def:g-PDA}
\end{defn}

\begin{defn} ({\bf$l$-cyclic $g$-regular  PDA} \cite{SR}): \label{def new PDA}
	In a $g-(K,F,Z,S)$ PDA P, if all the $Z$ stars in each column occur in consecutive rows and the position of stars in each column is obtained by cyclically shifting the previous column downwards by $l$ units, then it is called as \textit{$(l,g)-(K,F,Z,S)$ PDA}. 
	\label{def:l-g-PDA}
\end{defn}
	
	\section{Private Coded MADC Model}
	\label{dlmodel}
	
	In this section, we introduce a \textit{Partial Private Mapper Assignment MADC (PPMA-MADC) Model} that captures asymmetric knowledge about mapper-reducer connectivity.  
	In this model, mapper nodes are partitioned into \(K\) disjoint blocks of equal size \(\Lambda/K\), denoted \(\{\mathcal{M}_i: i \in [K]\}\).  
	Each reducer node \(R_k\) knows this global partition.  
	Block \(\mathcal{M}_k\) is the \textit{private block} of reducer \(R_k\); all other blocks \(\mathcal{M}_j\) (\(j \neq k\)) are \textit{public blocks} of reducer \(R_k\).  
	Accordingly:
	\begin{itemize}
		\item \(R_k\) is connected to {every} mapper in each public block (i.e., to all mappers in \((\bigcup_{i \in [K]} \mathcal{M}_i )\setminus \mathcal{M}_k\)), which is known to all reducers.
		\item \(R_k\) is connected to only a {subset} of the mappers in its private block \(\mathcal{M}_k\). This private connectivity pattern is {not known} to other reducers.
	\end{itemize}
	
	\noindent We study two specific private‑block connectivity patterns:
	
	\begin{enumerate}
		\item \textit{\(\alpha\)-connect PPMA-MADC Model:} Within its private block \(\mathcal{M}_k\), reducer \(R_k\) is connected to exactly \(\alpha\) mapper nodes, chosen arbitrarily. The specific connections are private to \(R_k\).
		
		\item \textit{\(\alpha\)-cyclic PPMA-MADC Model:} This model inherits all conditions of the \(\alpha\)-connect model, with the additional structure that the \(\alpha\) connected mappers in \(\mathcal{M}_k\) are consecutive under a cyclic wrap‑around ordering of the nodes in \(\mathcal{M}_k\). Other reducers are aware of this cyclic‑consecutive pattern but do not know the starting index (i.e., exactly which \(\alpha\) mappers are connected).
	\end{enumerate}
	
	The connectivity determines which file batches (and consequently which IVs) a reducer can access.  
	This work introduces privacy guarantees for the two connectivity patterns above. We further refer to these models as \textit{private  \(\alpha\)-connect PPMA-MADC model} and  \textit{private  \(\alpha\)-cyclic PPMA-MADC model}, respectively.
	For both these models,
	let \(\mathcal{R}_k \subseteq \mathcal{B}\) denote the set of batches accessible to reducer \(R_k\).  
	Given its assigned function \(\phi_{d_k}\) and accessible files \(\mathcal{R}_k\), reducer \(R_k\) generates and broadcasts a query \(\mathbf{y}_k\) to all other reducers. The privacy constraint that we put on both the models is defined below.
	
	\textit{Privacy Constraint:} No other reducer may learn the function index \(d_k\) assigned to \(R_k\).  
	Formally, for every \(k \in [K]\),
	\[
	I\big(\{d_j\}_{j \in [K]}; \{\mathbf{y}_j\}_{j \in [K]} \mid d_k, \mathcal{R}_k\big) = 0.
	\]
	
	After receiving all queries, each reducer \(R_k\) broadcasts a coded symbol \(\mathbf{X}_k\) that depends only on its accessible data \(\mathcal{R}_k\) and the received queries, i.e.,
$	H\big(\mathbf{X}_k \mid \mathcal{R}_k, \{\mathbf{y}_j\}_{j \in [K]}\big) = 0.$
	Finally, using the broadcast symbols \(\{\mathbf{X}_j\}_{j \in [K] \setminus k}\), each reducer \(R_k\) decodes all IVs required to compute its function \(\phi_{d_k}\).  
	Successful decoding requires $H\big(\phi_{d_k} \mid \mathcal{R}_k, d_k, \{\mathbf{y}_j\}_{j \in [K]}, \{\mathbf{X}_j\}_{j \in [K]}\big) = 0.$

	\section{Coding Scheme for Private \(\alpha\)-connect PPMA-MADC model}
	\label{main}

In this  section we first introduce an extended PDA constructed via \textbf{Algorithm \ref{algo1}} from a given base PDA.  
We then examine \(g\)-regular PDAs generated by \textbf{Algorithm \ref{algo2}}, which are based on the coded caching scheme of \cite{MAN}.  
By combining these two algorithms, we obtain a new family of extended PDAs described in \textbf{Construction \ref{CON1}}.  
These extended PDAs are used to provide coding scheme for  a private \(\alpha\)-connect PPMA-MADC scheme that supports \(K\) reducer nodes, \(Q={F \choose \alpha}\) output functions and $KF$ mapper nodes, for some integers $K, F$ and $\alpha$.  
The achievable computation-communication points for the scheme is characterized in \textbf{Theorem \ref{thm1}}, which summarizes the performance guarantees provided by Construction~\ref{CON1} in the private \(\alpha\)-connect setting. The correctness proofs for \textbf{Algorithm 1}, \textbf{Construction 1}, and \textbf{Theorem 1} are provided in Sections \ref{proof algo 1}, \ref{proof con 1} and \ref{proof thm1} respectively.

\subsection{Overview of {\bf Algorithm 1} and {\bf Construction \ref{CON1}}}
The construction in \textbf{Algorithm \ref{algo1}} builds on the extended PDA framework of \cite{sasi private tcom}.  
As the base PDA for that framework we take the array \(\mathbf{A} = [a_{i,k}]_{i,k \in [K]}\) whose entries are defined by \(a_{i,k} = *\) for \(i \neq k\) and \(a_{i,k} = 1\) for \(i = k\) (\(i,k \in [K]\)).  
This \(\mathbf{A}\) is a \((K,K,K-1,1)\) PDA.  
Using \(\mathbf{A}\) as the base PDA in the Algorithm 1 of \cite{sasi private tcom}, we present the adapted version as \textbf{Algorithm \ref{algo1}} in the present work.

	\begin{algorithm}
		\caption{Construction of a $(KQ,KF,(K-1)F + Z,S)$ PDA  ${\bf P}$ extended from  $(Q,F,Z,S)$   PDA ${\bf P}^{(1)}$ where $K >1$.}
		\label{algo1}
		${\bf P}$ is obtained  from ${\bf P}^{(1)}$ as follows:\begin{enumerate}
			\item Consider a $K\times K$ array ${\bf A} =(a_{i,k})$ where $a_{i,i}=1$ and $a_{i,k}=*$ for $i \neq k$, and $i,k\in [K]$.
			\item Replace the integer $1$ in ${\bf A} $ by the PDA ${\bf P}^{(1)}$.
			\item Replace each $*$ in ${\bf A}$ by a $F \times Q$ array denoted as $ {\bf X}$,  where all the entries are represented by the symbol $*$.
		\end{enumerate}
	\end{algorithm}

\begin{algorithm}
	\caption{${(\alpha+1)}\text{-}\left (F, {F \choose \alpha},{F-1 \choose \alpha-1}, {F \choose  \alpha+1} \right ) $ PDA $P_{F,\alpha}$ construction for some positive integers $F$ and $\alpha$ such that $\alpha \in [F-1] $.}
	\label{algo2}
	\begin{algorithmic}[1]
		\Procedure{{\bf 1}: }{}
		Arrange all subsets of size $ \alpha+1$ from $[F]$ in lexicographic order and for any subset $T'$ of size $\alpha+1$, define $y_{\alpha+1}(T')$ to be its order.
		\EndProcedure \textbf{ 1}
		\Procedure{{\bf 2}: }{}
		Obtain an array  $P_{F,\alpha}$ of size ${F \choose \alpha} \times {F }$.
		Denote the rows by the sets in $\{T\subset [F], |T| = \alpha \}$ and columns by the indices in $\{d: d\in [F]\}$. Define each entry $p_{T,d}$ corresponding to the row $T$ and the column $d$ as
		\begin{align}
		\label{Dk}
		p_{T,d} = \left\{
		\begin{array}{cc}
		*, &  \text{if } |T \cap d| \neq 0 \\
		y_{\alpha+1}(T \cup d), &  \text{if } |T \cap d| = 0 
		\end{array} \right\}.
		\end{align}
		\EndProcedure \textbf{ 2}
	\end{algorithmic}
\end{algorithm}

\begin{constr}
	\label{CON1}
	The following set of PDAs is derived using \textbf{Algorithm \ref{algo2}}, for positive integers \( F \) and \( \alpha \):
	
	\noindent {\small
	\begin{align}
	\Biggl \{   { P}_{\alpha} = (\alpha+1)\text{-}\left ({F}, {F\choose \alpha},{F-1 \choose \alpha-1},  {F \choose  \alpha+1} \right ) \nonumber \\ \text{ PDA}, \alpha \in [F-1] \Biggr \}.
	\end{align} }
We take transpose of each of the arrays in the above set to obtain the following set of PDAs.
	\noindent {\small
	\begin{align}
	\mathcal{P}^{(1)} =\Biggl \{   {\bf P}^{(1)}_{\alpha} = (\alpha+1)\text{-}\left ( {F\choose \alpha},F,\alpha,  {F \choose  \alpha+1} \right ) \nonumber \\ \text{ PDA}, \alpha \in [F-1] \Biggr \}.
	\end{align} }

	Now, for each \( \alpha \in [F-1] \), we consider the  PDA \({\bf P}^{(1)}_{\alpha} \in \mathcal{P}^{(1)}\).
	Using these arrays, we construct a new set of extended PDAs by applying \textbf{Algorithm 1}. The resulting set is given by:
	
	\noindent {\small
	\begin{align}
	\label{pda 1}
	\mathcal{P} =	\Biggl \{\left (  K{F \choose \alpha}, KF, (K-1)F+\alpha, {F \choose \alpha+1} \right ) \nonumber \\\text{ PDA}:  \alpha \in [F-1] \Biggr \}.
	\end{align}} 
\end{constr}
\subsection{Private coding scheme}
Using the PDAs constructed from {\bf Construction \ref{CON1}}, we now provide a coding scheme for a private \(\alpha\)-connect PPMA-MADC model.
	\begin{thm}
		\label{thm1}
		Consider a private \(\alpha\)-connect PPMA-MADC model with $KF$ mapper nodes, \(K\) reducer  nodes and \( {F \choose \alpha}\) output functions, for some positive integers $\alpha,K$ and $F$,  such that $\alpha \in [F-1]$.
		The PDAs obtained from {\bf Construction~1} corresponds to the model and achieves a  computation load  of $r = 1,$
		and  communication load  of
		\noindent {\small
			\begin{align}
			\label{com load}
			L_p= \frac{(F-\alpha)}{F(K-1)(\alpha +1)}. 
			\end{align}		}

	\end{thm}
\noindent Next, we illustrate Theorem 1 with the help of an example.
	\begin{example}
	\label{ex algo 3}
Consider a 2-connect PPMA-MADC system where we have a total of  \(9\) mapper nodes \(\{M_k^j : k,j \in [3]\}\), organized into \(3\) disjoint blocks of size \(3\): 
$\mathcal{M}_k = \{M_k^1, M_k^2, M_k^3\}, \quad \text{for each } k \in [3].$
There are \(3\) reducer nodes \(R_1, R_2, R_3\), each assigned to compute output function as follows: \(R_1\) computes \(\phi_1\), \(R_2\) computes \(\phi_2\), and \(R_3\) computes \(\phi_3\). 
There are  \(N = 9\) files \(\{W_i : i \in [9]\}\) and they are split into \(F = 9\) batches as follows: $
B_1^1 = \{W_1\},\; B_1^2 = \{W_2\},\; B_1^3 = \{W_3\},
B_2^1 = \{W_4\},\; B_2^2 = \{W_5\},\; B_2^3 = \{W_6\},
B_3^1 = \{W_7\},\; B_3^2 = \{W_8\},\; B_3^3 = \{W_9\}.$
Each mapper node \(M_k^j\) stores exactly one batch \(B_k^j\) for \(k,j \in [3]\).

	The  given PDA in  {\bf Construction \ref{CON1}} for $\alpha =2$ is defined as
${\small	{\bf P}^{(1)}  =
	\begin{bmatrix}
	\ast & \ast & 1 \\
	\ast & 1 & \ast \\
	1 & \ast & \ast
	\end{bmatrix}.}$ This PDA is obtained by putting $F=3$ and $\alpha =2$. Thus,
	 ${\bf P}^{(1)}$ is a $3$-$(3,3,2,1)$ PDA.
	The resulting extended array ${\bf P}_1$ obtained  using {\bf Construction  \ref{CON1}}  is
	\noindent {\small
		\begin{align}
		{\bf P}_1 =
		\begin{bmatrix}
		{\bf P}^{(1)} & {\bf X} & {\bf X} \\
		{\bf X} & {\bf P}^{(1)} & {\bf X} \\
		{\bf X} & {\bf X} & {\bf P}^{(1)}
		\end{bmatrix}, \quad
			{\bf X} = 
		\begin{bmatrix}
		\ast & \ast & \ast \\
		\ast & \ast & \ast \\
		\ast & \ast & \ast
		\end{bmatrix}.
		\label{pda ex 1}
		\end{align}}

\noindent	The constructed PDA \( {\bf P}_1 \) is a \((9,9,8,1)\) PDA.

	We consider a system with \(9\) effective reducer nodes \(\{R_k^j : j, k \in [3]\}\), consisting of the \(3\) real reducers and \(6\) virtual reducers.  
	The extended PDA \(\mathbf{P}_1\) is used to represent the connections, where each column corresponds to an effective reducer node and each row corresponds to a batch of files (equivalently, a mapper node).  
	A star \((\ast)\) in a given column indicates that the effective reducer node associated with that column is connected to the corresponding mapper node (i.e., the batch represented by that row is accessible at that effective reducer node).  
	The array \(\mathbf{P}_1\) is partitioned into \(3\) row blocks and \(3\) column blocks.  
	Within each column block \(k \in [K]\), the \(j\)-th column corresponds to the effective reducer node \(R_k^j\).  
	Exactly one column in the first block corresponds to the real reducer \(R_1\), while the others in that block are virtual; similarly, one column in the second block corresponds to \(R_2\), and likewise for the third block and \(R_3\).  
	On the row side, row block \(k \in [3]\) represents the mapper node block \(\mathcal{M}_k\), and the \(j\)-th row inside that block represents mapper node \(M_k^j\).  
	Since \(M_k^j\) stores batch \(B_k^j\), each row also uniquely represents the file batch \(B_k^j\).

 Each real reducer node \(R_k\) is connected to all mapper nodes in blocks \(\{\mathcal{M}_j : j \in [3] \setminus \{k\}\}\), but only to two mappers within its own block \(\mathcal{M}_k\).  
 The 2-element subsets of \([3] = \{1,2,3\}\) are ordered lexicographically: $
 \{1,2\},\; \{1,3\},\; \{2,3\}.$
 Denote by \(y_2(T)\) the index of the subset \(T\) in this order, so that $
 y_2(\{1,2\}) = 1, y_2(\{1,3\}) = 2, y_2(\{2,3\}) = 3.$
 Let the two mapper nodes in block \(\mathcal{M}_k\) to which \(R_k\) is connected be \(M_k^{t_1}\) and \(M_k^{t_2}\). Now, define
 $a_k = y_2(\{t_1,t_2\}).$
 The real reducer node \(R_k\) then corresponds to the \(a_k\)-th column within column block \(k\) of the extended PDA \(\mathbf{P}_1\).  
 Inside \(\mathbf{P}_1\), considering the column block $k$, the subarray \({\bf P}^{(1)}\) (located in row block \(k\) and column block \(k\)) has stars exactly in rows \(t_1\) and \(t_2\) of column \(a_k\), which matches the connectivity of \(R_k\) to those mapper nodes and hence the accessibility of the corresponding file batches.
 
 Thus, each real reducer \(R_k\) impersonates the effective reducer node \(R_k^{a_k}\).  
 For example, if \(a_1 = 1\), \(a_2 = 3\), and \(a_3 = 3\), then: $
 R_1$ { impersonates } $R_1^1$,
 $R_2$ { impersonates } $R_2^3$, and 
 $R_3$ { impersonates } $R_3^3.$
 The value \(a_k\) is private to reducer \(R_k\) and unknown to the other reducers.
Each real reducer node \(R_k\) has access to file batches as follows:
$\mathcal{R}_k =\bigl\{B_f^j : f \in [3] \setminus \{k\},\; j \in [3] \bigr\} \cup  \bigl\{B_k^j : j \in T,\; T \subseteq [3],\; y_2(T) = a_k,\; |T| = 2 \bigr\}.$
Consequently, \(R_k\)  can access the following IVs: $
 \{ v_{q,n} : q \in [3],\; W_n \in B,\; B \in \mathcal{R}_k \}.$
The real reducer \(R_1\) needs to compute \(\phi_1\) and is missing only the IV \(\{v_{1,3}\}\).  
Similarly, \(R_2\) (computing \(\phi_2\)) is missing \(\{v_{2,4}\}\), and \(R_3\) (computing \(\phi_3\)) is missing \(\{v_{3,7}\}\).
	
	Each real reducer node \(R_k\) generates a uniformly random permutation \(\mathbf{y}_k\) of the set \(\{1,2,3\}\) subject to the constraint that its \(a_k\)-th entry equals \(k\). This permutation is then broadcast to all other reducer nodes. As an example, if  
	$\mathbf{y}_1 = (1,3,2), \mathbf{y}_2 = (1,3,2), \mathbf{y}_3 = (2,1,3),$ 
	then the demands of the effective reducer nodes \(R_k^j\) are given by the sets \(D_k^j\) as follows:  $
	D_1^1 = \{v_{1,3}\},  D_1^2 = \{v_{3,2}\},  D_1^3 = \{v_{2,1}\}, 
	D_2^1 = \{v_{1,6}\},  D_2^2 = \{v_{3,5}\},  D_2^3 = \{v_{2,4}\}, 
	D_3^1 = \{v_{2,9}\},  D_3^2 = \{v_{1,8}\},  D_3^3 = \{v_{3,7}\}.$
	This is since, ${\bf y}_k[j]$ represents the output function index assigned to the effective reducer node $R_k^j,$ for $k,j \in [3]$.
	
	The integer entries in the extended PDA \(\mathbf{P}_1\) represent these demands. For instance, the \(1\) appearing in the first column and third row of \(\mathbf{P}_1\) indicates that effective node \(R_1^1\) demands the IV associated with the file contained in batch \(B_1^3\) and the output function \(\phi_1\) (the function assigned to \(R_1^1\)). Since \(B_1^3 = \{W_3\}\), this entry corresponds to the IV \(v_{1,3}\).
	
The IVs are split into two equal halves, with superscripts assigned according to the following rule: for each integer entry in the PDA, the superscripts are taken as the indices of the other column blocks (excluding the block containing the entry). For example, the demand corresponding to the integer \(1\) in the first column (namely \(v_{1,3}\)) is divided into two equal parts, labeled with superscripts \(2\) and \(3\) (these are precisely the column-block indices different from the block to which the column belongs). Thus, we have $
		v_{1,3} = \{v_{1,3}^2, v_{1,3}^3\},  v_{3,2} = \{v_{3,2}^2, v_{3,2}^3\},  v_{2,1} = \{v_{2,1}^2, v_{2,1}^3\} ,
	v_{1,6} = \{v_{1,6}^1, v_{1,6}^3\},  v_{3,5} = \{v_{3,5}^1, v_{3,5}^3\},  v_{2,4} = \{v_{2,4}^1, v_{2,4}^3\}, 
		v_{2,9} = \{v_{2,9}^1, v_{2,9}^2\},  v_{1,8} = \{v_{1,8}^1, v_{1,8}^2\},  v_{3,7} = \{v_{3,7}^1, v_{3,7}^2\}.$
	Each real reducer node transmits a coded message constructed as follows.  
	\(R_1\) takes the portions of the demanded IVs that correspond to the symbol \(1\) appearing in columns of column blocks \(2\) and \(3\), provided those portions carry the superscript \(1\).  
	It then XORs these parts together and broadcasts to the other real reducer nodes.  
	Similarly, \(R_2\) takes the parts of demands corresponding to the symbol \(1\) in columns of column blocks \(1\) and \(3\) that have superscript \(2\), XORs them, and transmits the result.  
	Real node \(R_3\) proceeds in the same way, using demands from column blocks \(1\) and \(2\) that carry superscript \(3\).
	The transmitted coded symbols are
$	X_1^1 = v_{2,9}^1 \oplus v_{1,8}^1 \oplus v_{3,7}^1 \oplus v_{1,6}^1 \oplus v_{3,5}^1 \oplus v_{2,4}^1,
	X_2^1 = v_{2,9}^2 \oplus v_{1,8}^2 \oplus v_{3,7}^2 \oplus v_{1,3}^2 \oplus v_{3,2}^2 \oplus v_{2,1}^2,
	X_3^1 = v_{1,6}^3 \oplus v_{3,5}^3 \oplus v_{2,4}^3 \oplus v_{1,3}^3 \oplus v_{3,2}^3 \oplus v_{2,1}^3$.
Real reducer node \(R_1\) retrieves the part \(v_{1,3}^2\) from the transmission \(X_2^1\) and the part \(v_{1,3}^3\) from \(X_3^1\).  
In a similar manner, every real reducer node obtains the  IVs required to compute its assigned output function.
Each mapper node stores one batch of files, giving a computation load of \(r = 1\).  
A total of \(3\) coded symbols are broadcast by the reducer nodes, each of size \(\beta/2\).  
Thus, the resulting communication load is  
$
L_p = \frac{3 \cdot (\beta/2)}{3 \cdot 9 \cdot \alpha} = \frac{1}{18}.
$
	
\end{example}
\section{Coding Scheme for Private \(\alpha\)-cyclic  PPMA-MADC model}
\label{main2}
In this section we consider \(\alpha\)-cyclic  PPMA-MADC model and provide private coding scheme for this model when $\alpha < \frac{Q}{2}$ and $Q+ \alpha $ is even. We first construct $(1,2)$-regular PDAs using {\bf Algorithm 3}.
We then combine  {\bf Algorithms 1} and {\bf 3} to obtain a new family of extended PDAs, formally described in \textbf{Construction \ref{CON2}}. 
We use these extended PDAs produced by \textbf{Construction \ref{CON2}} to derive coding scheme for a private $\alpha$-cyclic PPMA-MADC model serving $K$ reducer nodes, $Q $ output functions and $KQ$ mapper nodes. The achievable computation-communication points is characterized in \textbf{Theorem \ref{thm2}}. The proofs of correctness of {\bf Algorithm \ref{alg:structured-array}, {\bf Construction \ref{CON2}}} and {\bf Theorem \ref{thm2}} are provided in Sections \ref{proof algo 3}, \ref{proof con 2} and \ref{proof thm2} respectively.

	\begin{algorithm}
	\caption{$(1,2)-\left(Q, Q, \alpha, \frac{Q(Q-\alpha)}{2} \right)$ PDA $P$ Construction}
	\label{alg:structured-array}
	\begin{algorithmic}[1]
		\Require $Q \geq 2$, $\alpha$ satisfying $0 < \alpha < \frac{Q}{2}$, where both $Q$ and $\alpha$ are even or both are odd
		
		\State \textbf{Step 1: Initial Setup}
		\State $m \gets \frac{Q + \alpha}{2}$
		\State Create $Q \times Q$ array $P=[p_{i,j}]_{i,j \in [Q]},$ initialized with `*' as entries
		
		\State \textbf{Step 2: Fill rows indexed from  $\alpha+1$ \textbf{to} $m$}
		\For{$r = \alpha+1$ \textbf{to} $m$}
		\State $\text{start\_num} \gets (r - (\alpha +1)) \times Q + 1$
		\For{$c = 1$ \textbf{to} $Q$}
		\State $p_{r,c} \gets \text{start\_num} + (c - 1)$
		\EndFor
		\EndFor
		
		\State \textbf{Step 3: Fill rows indexed from  $m+1$ \textbf{to} $Q$}
		\For{$j = 1$ \textbf{to} $(m - \alpha)$}
		\State $\text{source\_row} \gets m - j + 1$
		\State $\text{target\_row} \gets m + j$
		\State $\text{shift\_amount} \gets m - j$
		\For{$c = 1$ \textbf{to} $Q$}
		\State $\text{col} \gets \left [c + \text{shift\_amount} \right ]_Q$
		\State $p_{\text{target\_row},\text{col}} \gets p_{\text{source\_row},c}$
		\EndFor
		\EndFor
		
		\State \textbf{Step 4: Apply Column Shifts}
		\State Create temporary array $A=[a_{i,j}]_{i,j \in [Q]}$ as copy of $P$
		\For{$col = 1$ \textbf{to} $Q$}
		\State $\text{shift} \gets col - 1$
		\For{$row = 1$ \textbf{to} $Q$}
		\State $\text{new\_row} \gets \left [row + \text{shift}  \right ]_Q$
		\State $a_{\text{new\_row},col} \gets a_{row,col}$
		\EndFor
		\EndFor
		\State $P \gets A$
		
		\State \Return $P$
	\end{algorithmic}
\end{algorithm}

\subsection{Construction of $(1,2)$-regular  PDAs using {\bf Algorithm 3}}
\label{algo 3 overview}

	The algorithm constructs a $Q \times Q$ array with entries as stars and integers. The construction proceeds through four systematic steps, each building upon the previous one to create the final array with the desired combinatorial property.

\noindent {\textbf{Step 1}}: The construction begins by establishing key parameters. Given an  integer $Q \geq 2$ and an  integer $\alpha$ satisfying $\alpha < \frac{Q}{2}$, where both $Q$ and $\alpha$ are even or both are odd, we define $m = \frac{Q + \alpha}{2}$ (since both $Q$ and $\alpha$ are even or both are odd, $Q+\alpha$ is even).  The entire $Q \times Q$ array is initially populated with stars (`*'), which will be selectively replaced by numeric values in subsequent steps. This initialization ensures that any positions not explicitly assigned numeric values will maintain their entries as stars  throughout the construction.

\noindent {\textbf{Step 2}}: The second step fills rows $\alpha +1$ through $m$ with consecutive integers in blocks of size $Q$. Specifically, for each row $r$ in the range $\alpha +1 \leq r \leq m$, we assign the values $(r - (\alpha +1)) \times Q + 1$ through $(r - (\alpha +1)) \times Q + Q$ to columns $1$ through $Q$, respectively. The entries are filled row-wise with sequential integers. Row $\alpha +1$ contains the integers 1 through Q, row $\alpha +2$ contains Q+1 through 2Q, and this pattern continues for subsequent rows. This creates disjoint blocks of consecutive integers where each integer appears exactly once in this section of the array. The entries of  first $\alpha$ rows remain as stars.

\noindent {\textbf{Step 3}}: The third step creates the lower portion of the array (rows $m+1$ through $Q$) by applying systematic right cyclic shifts to the numeric rows from the previous step. For each $j = 1, 2, \ldots, (m - \alpha)$, we take the row indexed by  $m - j + 1$ and copy it to row indexed by $m + j$ after applying a right cyclic shift of $m - j$ positions. This creates a symmetric relationship where each numeric row $r_1$ in the range $[\alpha +1, m]$ appears exactly twice in the array: once in its original form at row $r_1$, and once in a shifted form at row $r_1' = 2m - r + 1$. The shift amounts are carefully chosen to decrease as we move downward through the array, creating a structured pattern of repetitions. Hence there are $S=Q(m-\alpha)$ integers appearing in this array and each integer appears $g=2$ times in the array.

\noindent {\textbf{Step 4}}: The final step applies independent cyclic shifts to each column of the array. For column $j$ (where $ j \in[Q] $), we shift all entries downward by $j-1$ positions, with cyclic wrap-around. 
\begin{example}
			This example provides a detailed walkthrough of {\bf Algorithm 3} for constructing a $6 \times 6$ array with parameters $\alpha=2$ and $Q=6$. The algorithm builds a PDA with specific combinatorial properties through four systematic steps.
		
		Step 1: The construction begins by computing the key parameter $m = \frac{Q + \alpha}{2} = \frac{6 + 2}{2} = 4$. A $6 \times 6$ array $A$ is created and initialized entirely with stars (`*'), which serve as placeholder entries. 
		\[A=
		\begin{bmatrix}
		* & * & * & * & * & * \\
		* & * & * & * & * & * \\
		* & * & * & * & * & * \\
		* & * & * & * & * & * \\
		* & * & * & * & * & * \\
		* & * & * & * & * & *
		\end{bmatrix}
		\]
		
		Step 2: 
		The second step fills rows indexed from $\alpha+1 = 3$ to $m = 4$ with consecutive integers in blocks of size $Q = 6$. For row $r = 3$, the starting number is calculated as $(r - (\alpha + 1)) \times Q + 1 = (3 - 3) \times 6 + 1 = 1$, resulting in the sequence $[1, 2, 3, 4, 5, 6]$ placed in columns 1 through 6. For row $r = 4$, the starting number is $(4 - 3) \times 6 + 1 = 7$, producing the sequence $[7, 8, 9, 10, 11, 12]$. This creates two complete rows of consecutive integers where each integer from 1 to 12 appears exactly once in this section of the array. The first $\alpha = 2$ rows remain filled with stars at this stage.
		\[
		A = 
		\begin{bmatrix}
		* & * & * & * & * & * \\
		* & * & * & * & * & * \\
		1 & 2 & 3 & 4 & 5 & 6 \\
		7 & 8 & 9 & 10 & 11 & 12 \\
		* & * & * & * & * & * \\
		* & * & * & * & * & *
		\end{bmatrix}
		\]
		Step 3: 
		The third step creates the lower portion of the array (rows 5 and 6) by applying systematic right cyclic shifts to the numeric rows from Step 2. For $j = 1$, we take row $m - j + 1 = 4$ (containing $[7, 8, 9, 10, 11, 12]$) and copy it to row $m + j = 5$ after applying a right cyclic shift of $m - j = 3$ positions, resulting in $[10, 11, 12, 7, 8, 9]$. For $j = 2$, row $m - j + 1 = 3$ (containing $[1, 2, 3, 4, 5, 6]$) is copied to row $m + j = 6$ with a right cyclic shift of $m - j = 2$ positions, yielding $[5, 6, 1, 2, 3, 4]$. This creates a symmetric structure where each numeric value from rows 3 and 4 appears exactly twice in the array—once in its original position and once in a shifted position in rows 5 or 6. The shift amounts decrease systematically as we move downward through the array.
		\[
		A= 
		\begin{bmatrix}
		* & * & * & * & * & * \\
		* & * & * & * & * & * \\
		1 & 2 & 3 & 4 & 5 & 6 \\
		7 & 8 & 9 & 10 & 11 & 12 \\
		10 & 11 & 12 & 7 & 8 & 9 \\
		5 & 6 & 1 & 2 & 3 & 4
		\end{bmatrix}
		\]
		
		Step 4: 
		The final step applies independent cyclic shifts to each column. For column $j$, all entries are shifted downward by $j-1$ positions with cyclic wrap-around. Column 1 (shift=0) remains unchanged. Column 2 (shift=1) shifts its entries $[*, *, 2, 8, 11, 6]^{T}$ down by one position to become $[6, *, *, 2, 8, 11]^{T}$. Column 3 (shift=2) transforms $[*, *, 3, 9, 12, 1]^{T}$ into $[12, 1, *, *, 3, 9]^{T}$. Column 4 (shift=3) converts $[*, *, 4, 10, 7, 2]^{T}$ to $[10, 7, 2, *, *, 4]^{T}$. Column 5 (shift=4) changes $[*, *, 5, 11, 8, 3]^{T}$ to $[5, 11, 8, 3, *, *]^{T}$. Finally, column 6 (shift=5) transforms $[*, *, 6, 12, 9, 4]^{T}$ into $[*,6, 12, 9, 4,  *]^{T}$.
		
		The resulting $6 \times 6$ PDA ${\bf P}^{(2)}$ with $\alpha=2$ and $Q=6$ is:
		
		\begin{align}
		\label{pda consecutive}
		{\bf P}^{(2)}= 
		\begin{bmatrix}
		* & 6 & 12 & 10 & 5 & * \\
		* & * & 1 & 7 & 11 & 6 \\
		1 & * & * & 2 & 8 & 12 \\
		7 & 2 & * & * & 3 & 9 \\
		10 & 8 & 3 & * & * & 4 \\
		5 & 11 & 9 & 4 & * & *
		\end{bmatrix}
	\end{align}
		
It can be verified that it is a consecutive and $(1,2)-(6,6,2,12)$ PDA.

\end{example}
\begin{constr}
	\label{CON2}
	The following set of PDAs is derived using \textbf{Algorithm 3}, for positive integers \( Q \) and \( \alpha \) such that $\alpha<\frac{Q}{2}$ and $Q+\alpha$ is even:
	
	\noindent {\small 
	\begin{align}
	\mathcal{P}^{(2)} =\Biggl \{   {\bf P}^{(2)}_{\alpha} =  (1,2)\text{-}\left ({Q}, {Q},\alpha,\frac{Q(Q-\alpha)}{2} \right ) \text{ PDA}:   \nonumber \\\alpha \in \left[1, \left\lceil \frac{Q}{2}\right \rceil \right ), Q+\alpha \text{ even} \Biggr \}.
	\end{align} }
	
\noindent
	Now, for each $\alpha \in \left[1, \left\lceil \frac{Q}{2}\right \rceil \right )$ such that $ Q+\alpha \text{ is even} $, we consider the  PDA \({\bf P}^{(2)}_{\alpha} \in \mathcal{P}^{(2)}\).
	Using these arrays, we construct a new set of extended PDAs by applying \textbf{Algorithm 1}. The resulting set is given by:
	
	\noindent {\small 
	\begin{align}
	\label{pda 2}
	\mathcal{P} =	\Biggl \{\left ( KQ, KQ,(K-1)Q + \alpha,  \frac{Q(Q-\alpha)}{2}  \right ) \text{ PDA}: \nonumber \\ \alpha \in \left[1, \left\lceil \frac{Q}{2}\right \rceil \right ), Q+\alpha \text{ even}\Biggr \}.
	\end{align}} 
\end{constr}
\subsection{Private Coding Scheme for \(\alpha\)-cyclic  PPMA-MADC model}
Using the PDAs constructed from {\bf Construction \ref{CON2}}, we now provide the private coding scheme for a private \(\alpha\)-cyclic PPMA-MADC model.

\begin{thm}
	\label{thm2}
Consider a private \(\alpha\)-cyclic  PPMA-MADC model with $KQ$ mapper nodes, \(K\) reducer nodes and \(Q\) output functions,  for some positive integers $ K,$ and $ Q$  such that 
$\alpha < \frac{Q}{2}$  and $Q + \alpha $  even. 
The PDAs obtained from {\bf Construction \ref{CON2}} corresponds to the model and achieves a  computation load  of $r = 1,$
and  communication load  of
\begin{align}
L_p &= \frac{\,(Q-\alpha)}{2\,Q\,\,(K-1)} .
\end{align}
\end{thm}
	Now, we illustrate Theorem \ref{thm2} using a simple example.
	\begin{example}
		\label{ex pda 1}
		In a \(2\)-cyclic PPMA-MADC system, we have \(N = 36\) files: \(\{W_i^j : i, j \in [6]\}\), which are partitioned into \(F = 36\) batches \(B_i^j = \{W_i^j\}\) for \(i,j \in [6]\).  
		The system consists of \(\Lambda = 36\) mapper nodes $\{M_k^j: k,j \in [6]\}$ organized into disjoint mapper node blocks of size 6, $\M_k=\{M_k^j: j \in {6}\}$, for $k\in [6]$. There are  \(K = 6\) reducer nodes $\{R_k: k\in [6]\}$, and \(Q = 6\) output functions $\{\phi_i, i \in [6]\}$.  
		Each mapper node \(M_i^j\) stores exactly one batch \(B_i^j\) of files, for \(i,j \in [6]\).  
		Each reducer node is assigned a single output function to compute. Assume that reducer node \(R_k\) computes \(\phi_k\), for  \(k \in [6]\).
		
		The PDA for this example is defined as ${\bf P}^{(2)}$ in (\ref{pda consecutive}).		
		The extended PDA ${\bf P}_2$ is constructed from this PDA using {\bf Algorithm 1} as follows:

	\begin{align}
	{\bf P}_2 = 
	\begin{pmatrix}
	{\bf P}^{(2)} & {\bf X} & {\bf X} & {\bf X} & {\bf X} & {\bf X} \\
	{\bf X} & {\bf P}^{(2)}  & {\bf X} & {\bf X} & {\bf X} & {\bf X} \\
	{\bf X} & {\bf X} & {\bf P}^{(2)}  & {\bf X} & {\bf X} & {\bf X} \\
	{\bf X} & {\bf X} & {\bf X} & {\bf P}^{(2)}  & {\bf X} & {\bf X} \\
	{\bf X} & {\bf X} & {\bf X} & {\bf X} & {\bf P}^{(2)}  & {\bf X} \\
	{\bf X} & {\bf X} & {\bf X} & {\bf X} & {\bf X} & {\bf P}^{(2)} 
	\end{pmatrix}
	\end{align}
	where ${\bf X}=
				\begin{bmatrix}
				* & * & * & * & * & * \\
				* & * & * & * & * & * \\
				* & * & * & * & * & * \\
				* & * & * & * & * & * \\
				* & * & * & * & * & * \\
				* & * & * & * & * & *
				\end{bmatrix}
				$.

		We introduce \(30\) virtual reducer nodes alongside the \(6\) real reducer nodes, resulting in a total of \(36\) effective reducer nodes \(\{R_k^j : k, j \in [6]\}\) (with \(30\) virtual and \(6\) real).  
		In the extended PDA \(\mathbf{P}_2\), each column corresponds to an effective reducer node, and each row corresponds to a batch of files (equivalently, a mapper node). An asterisk \((\ast)\) in a given column indicates that the batch associated with that row is accessible at that effective reducer node.   
		The array \(\mathbf{P}_2\) is structured as a block matrix with \(6\) row blocks and \(6\) column blocks. Within each column block \(k \in [6]\), the \(j\)-th column corresponds to the effective reducer node \(R_k^j\).  Also, in each column block \(k\), exactly one column corresponds to the real reducer node \(R_k\); the remaining columns in that block represent virtual reducers.  
		Similarly, the row blocks of \(\mathbf{P}_2\) correspond to mapper node blocks: the first row block corresponds to mapper block \(\mathcal{M}_1\), the second row block to \(\mathcal{M}_2\), and so on.

		Each real reducer node \(R_k\) (where \(k \in [6]\)) impersonates one of the columns from column block \(k\) in the extended PDA, depending on the specific \(\alpha\) consecutive mapper nodes it is connected to within its own mapper node block \(\mathcal{M}_k\).  
		If \(R_k\) is connected to mapper nodes in the circular interval $\{M_k^j: j \in [a_k, a_k+1]_6\}$ inside \(\mathcal{M}_k\), then it impersonates the \(a_k\)-th column of column block \(k\).  
		As an example, if \(a_k = k\) for all \(k \in [6]\), then real reducer node \(R_k\) impersonates effective node \(R_k^k\).  
		The value \(a_k\) is private to \(R_k\) and unknown to the other reducers.
		
		Notice that in the PDA \(\mathbf{P}_2\), all entries of column block \(k\) in row blocks \(j \in [6] \setminus \{k\}\) are filled with the symbol \(\mathbf{X}\), indicating that reducer node \(R_k\) is connected to every mapper node in blocks \(\mathcal{M}_j\) for \(j \neq k\).  
		Within row block \(k\) and column block \(k\), the subarray is the consecutive \(1\)-cyclic PDA \({\bf P}^{(2)}\).  
		In \({\bf P}^{(2)}\), each column contains \(2\) consecutive star entries: the first column has stars in rows \(1\) and \(2\), the second column in rows \(2\) and \(3\), and so on cyclically.  
		Consequently, if \(R_k\) is connected to mapper nodes in the circular interval \([a_k,\; a_k+1]_6\) within \(\mathcal{M}_k\), it possesses access to batches \(B_k^{a_k}\) and \(B_k^{(a_k +1) \bmod 6}\).  
		This access pattern corresponds precisely to column \(a_k\) within column block \(k\) of the PDA, confirming that the PDA  models the problem.
		
		Each real reducer node has access to the following collection of batches:
		
		\noindent {\small 
		\begin{align}
		\mathcal{R}_k
		= \{ B_j^f : j \in [6]\setminus\{k\},\; f \in [6] \}
		\;\cup\;
		\{ B_k^f : f \in [a_k, a_k+1]_6 \}.
		\end{align}}
		
	\noindent	Consequently, real reducer node \(R_k\) is able to compute the IVs
		\begin{align}
		\{ v_{q,n} : q \in [6],\; n \in B,\; B \in \mathcal{R}_k \}.
		\end{align}
		
		Each real reducer node \(R_k\) is responsible for computing the function \(\phi_k\).
		However, some of the required IVs are not locally available at \(R_k\).
		In particular, for real reducer node \(R_1\), the missing IVs are
		\begin{align}
		\{ v_{1,n} : W_n \in B_1^3 \cup B_1^4 \cup B_1^5 \cup B_1^6 \}.
		\end{align}
		
		More generally, for real reducer node \(R_k\), the missing IVs are given by
		\begin{align}
		\{ v_{k,n} : W_n \in B_k^f,\; f \in [a_k+2, a_k+5]_6 \}.
		\end{align}

Each real reducer node \(R_k\) generates a uniformly random permutation \(\mathbf{y}_k\) of the set \([6]\), subject to the condition that its \(a_k\)-th entry equals \(d_k=k\), and broadcasts it to all other reducers. For example,
$\mathbf{y}_1 = \mathbf{y}_2 = \mathbf{y}_3 = 
\mathbf{y}_4 = \mathbf{y}_5 = \mathbf{y}_6 = (1,2,3,4,5,6).$

For each \(f\) in the circular interval \([a_k+2,\; a_k+5]_6\), we aggregate the  IVs required by the effective reducer node \(R_k^{a_k}\) (which is impersonated by the real node \(R_k\)) that can be computed from the files in batch \(B_k^f\). This aggregated symbol is
\[
\U_{d_k, B_k^f} = \bigl(v_{d_k,n} : W_n \in B_k^f\bigr) \in \mathbb{F}_{2^{\eta \beta}},
\]
where \(\phi_{d_k} = \phi_{\mathbf{y}_k[a_k]}\) is the output function that \(R_k\) must compute.

The symbol \(\U_{d_k, B_k^f}\) is then partitioned into \(5\) equally sized packets:
\[
\U_{d_k, B_k^f} = \bigl\{\U_{d_k, B_k^f}^{\,j} : j \in [6] \setminus \{k\}\bigr\}.
\]

The integer entries in the PDA \(\mathbf{P}_2\) denotes the demands of the effective nodes. For instance, a \(1\) appearing in the first column and third row of \(\mathbf{P}_2\) indicates that the corresponding effective node requests the aggregated IV set associated with batch \(B_1^3\) and output function \(\phi_1\); hence that entry represents \(\U_{1, B_1^3}\).

More generally, for each \(k \in [6]\) and each \(f\) in the circular interval \([j+2,\; j+5]_6\) (with \(j \in [6]\)), we form the aggregated symbol \(\U_{\mathbf{y}_k[j], B_j^f}\) containing all IVs that effective node \(R_k^j\) is assumed to compute from batch \(B_j^f\). This symbol is likewise split into \(5\) equal‑sized packets, each labeled with a distinct superscript from the set \([6] \setminus \{k\}\).

The superscript assignment rule is that the demand corresponding to an integer entry in a given column block takes as superscripts the indices of the other five column blocks. For example, the demand associated with the integer \(1\) in the first column (namely \(\U_{1, B_1^3}\)) is divided into \(5\) equal parts with superscripts \(2,3,4,5,6\), which are precisely the column block indices different from the block containing that column.

Let $\mathcal{J}_t^k = \{(j,f,i) : {\bf P}^{(2)}_{f,i} = t,\; j \in [6] \setminus \{k\},\; f,i \in [6]\}.$
Now, each real reducer node constructs coded transmissions as follows:  
Real node \(R_1\) collects the packet parts of the demands that correspond to the integer \(1\) appearing in columns of all column blocks except block~1, XORs them, and broadcasts the result to the other real nodes.  
The same procedure is repeated for every integer \(t \in [12]\) present in the PDA (since there are \(12\) distinct integers in total).  
Similarly, \(R_2\) processes the packet parts from all column blocks except block~2, and so on for every \(k \in [6]\).

Formally, for each \(k \in [6]\) and for each \(t \in [12]\), real reducer node \(R_k\) produces the coded symbol
\begin{align}
X_k^t = \bigoplus_{(j,f,i) \in \mathcal{J}_t^k} \U_{\mathbf{y}_k[i], B_k^{f}}^{\,k},
\end{align}
and multicasts the sequence $\mathbf{X}_k = \{X_k^t : t \in [12]\}.$
Real node \(R_1\) can thus recover the whole aggregated symbol \(\U_{1,B_1^3}\) by obtaining its constituent packets \(\U_{1,B_1^3}^{\,j}\) from the transmissions \(X_j^1\) for every \(j \in [6] \setminus \{1\}\).  
All other real nodes likewise retrieve the data required to compute their assigned output functions.

In this scheme, each mapper node stores exactly one batch of files, so the computation load is \(r = 1\).  
A total of \(12\) coded symbols are broadcast, each of size \(\beta/5\).  
Therefore, the communication load is
$L = \frac{12 \cdot (\beta/5)}{6 \cdot 6 \cdot \beta} = \frac{2}{30}.$
		
	\end{example}
	\section{Proof of {\bf Algorithm 1}}
	\label{proof algo 1}
	We have seen that \textbf{Algorithm \ref{algo1}} is based on the extended PDA framework of \cite{sasi private tcom}.  In this case we fix the first PDA as 
 the array \(\mathbf{A} = [a_{i,k}]_{i,k \in [K]}\) whose entries are defined by \(a_{i,k} = *\) for \(i \neq k\) and \(a_{i,k} = 1\) for \(i = k\) (\(i,k \in [K]\)).  
	This \(\mathbf{A}\) is a \((K,K,K-1,1)\) PDA.  
	Using \(\mathbf{A}\) as the first PDA and PDA ${\bf P}^{(1)}$ as the second PDA in the Algorithm 1 of \cite{sasi private tcom}, the resulting PDA is the PDA  {\bf P} obtained in {\bf Algorithm 1}. The proof of construction follows directly from the proof of Algorithm 1 in \cite{sasi private tcom}.
	
	\section{Proof of {\bf Construction 1}}
	\label{proof con 1}
	Let $F$ be a positive integer and $\alpha \in [F-1]$. In {\bf Algorithm 2}, $P_{F,\alpha}$ is a $\binom{F}{\alpha} \times F$ array whose rows are indexed by $\alpha$-subsets $T$ of $[F]$ and columns by elements $d \in [F]$. Each entry $p_{T,d}$ is $*$ if $d \in T$; otherwise it is the lexicographic index of the $(\alpha+1)$-subset $T \cup \{d\}$. This yields a PDA with parameters  
	\[
	(\alpha +1)-\left(F,\ \binom{F}{\alpha},\ \binom{F-1}{\alpha-1},\ \binom{F}{\alpha+1}\right).
	\]  
	
	The transpose $P_{F,\alpha}^T$ is an $F \times \binom{F}{\alpha}$ array, with rows indexed by $d \in [F]$ and columns indexed by $\alpha$-subsets $T$. Its entry at $(d, T)$ is simply $p_{T,d}$ from the original array. To show that $P_{F,\alpha}^T$ is also a PDA, we verify the three PDA conditions.
	
	\noindent
	\textit{A1 (Star count per column):} In a fixed column $T$ of $P^T$, the star appears precisely when $d \in T$. Since $|T| = \alpha$, there are exactly $\alpha$ stars in that column. Thus $Z' = \alpha$.
	
	\noindent
	\textit{A2' (Each integer appears $\alpha +1$ times):} Each integer $s$ corresponds uniquely to an $(\alpha+1)$-subset $Y$. In the original PDA, $s$ appears at positions $(T, d)$ where $T = Y \setminus \{d\}$ and $d \in Y$. In the transpose, $s$ therefore appears at positions $(d, T)$ for each $d \in Y$ with $T = Y \setminus \{d\}$. Since $Y$ has $\alpha+1$ elements, $s$ occurs at least once (in fact, $\alpha+1$ times) in $P^T$.
	
	\noindent
	\textit{A3 (Integer condition):} Suppose the same integer $s$ appears in two distinct positions $(d_1, T_1)$ and $(d_2, T_2)$ of $P^T$. Then $s$ corresponds to an $(\alpha+1)$-subset $Y$ such that $Y = T_1 \cup \{d_1\} = T_2 \cup \{d_2\}$. From this, $T_1 \neq T_2$ and $d_1 \neq d_2$, so the two positions lie in distinct rows and distinct columns. Now consider the $2\times 2$ subarray of $P^T$ formed by rows $d_1, d_2$ and columns $T_1, T_2$. The entry $(d_1, T_2)$ equals $p_{T_2, d_1}$ in the original array. Since $d_1 \in Y$ and $T_2 = Y \setminus \{d_2\}$ with $d_1 \neq d_2$, we have $d_1 \in T_2$; hence $p_{T_2, d_1} = *$. Similarly, $d_2 \in T_1$, so $p_{T_1, d_2} = *$. Thus the subarray is either  
	\[
	\begin{pmatrix}
	s & * \\
	* & s
	\end{pmatrix}
	\quad\text{or}\quad
	\begin{pmatrix}
	* & s \\
	s & *
	\end{pmatrix},
	\]  
	fulfilling condition A3.
	
	Therefore $P_{F,\alpha}^T$ satisfies all PDA conditions with parameters  
	\[
	(\alpha +1)-\left(\binom{F}{\alpha},\ F,\ \alpha,\ \binom{F}{\alpha+1}\right),
	\]  
	which is precisely the set ${\bf P}^{(1)}_{\alpha}$ described in the construction. Hence for each $\alpha \in [F-1]$, the PDA \({\bf P}^{(1)}_{\alpha}\)  is a $(\alpha+1)$-$\left (  {F \choose \alpha},F,\alpha, {F \choose \alpha+1} \right )$ PDA.
	
	Each array \({\bf P} \in \mathcal{P}\) is structured as a block array consisting of \(K\) row blocks indexed by \(\left [{K} \right ]\) and \(K\) column blocks indexed by \([K]\). The block entries are denoted as \(\{{\bf p}_{f,k} : f, k \in [K]\}\). These blocks include the arrays \({\bf X}\) and \({\bf P}_{\alpha}^{(1)} \), each of size \(F \times {F \choose \alpha}  \). Consequently, the overall size of \({\bf P}\) is \(KF \times K{F \choose \alpha } \). 
	
	Each column in the PDA \({\bf P}^{(1)}_{\alpha}\) contains exactly $\alpha$ stars, and  the array \({\bf X}\) contains exactly $F$ stars per column. Thus, each column in \({\bf P}\) contains $\alpha +(K-1){F }$ stars, satisfying condition \(A1\) of Definition \ref{def:PDA} with \(Z = \alpha +(K-1){F }\). 
The total number of integers in \({\bf P}\) is \({F \choose \alpha+1}\). This satisfies condition \(A2\) of Definition \ref{def:PDA}.
	
	From \textbf{Construction \ref{CON1}}, for two distinct block entries \({\bf p}_{f_1,k_1}\) and \({\bf p}_{f_2,k_2}\), \({\bf p}_{f_1,k_1} = {\bf p}_{f_2,k_2} = {\bf P}_{\alpha}^{(1)}\)  only if \(f_1 \neq f_2\) and \(k_1 \neq k_2\). These entries belong to distinct row and column blocks. Furthermore, \({\bf p}_{f_1,k_2} = {\bf p}_{f_2,k_1} = {\bf X}\), meaning the corresponding \(2 \times 2\) sub-block array formed by row blocks \(f_1, f_2\) and column blocks \(k_1, k_2\) must have one of the following forms:
	\[
	\begin{pmatrix}
	{ {\bf P}^{(1)}_{\alpha}} & {\bf X} \\
	{\bf X} & 	{ {\bf P}^{(1)}_{\alpha}}
	\end{pmatrix}
	\quad \text{or} \quad
	\begin{pmatrix}
	{\bf X} & 	{ {\bf P}^{(1)}_{\alpha}} \\
		{ {\bf P}^{(1)}_{\alpha}} & {\bf X}
	\end{pmatrix}.
	\]

	In conclusion, the array \({\bf P}\) is a $\left ( K{F \choose \alpha},KF,\alpha +(K-1){F }, {F \choose \alpha+1} \right )$  PDA.
	
	\section{Proof of Theorem \ref{thm1}}
	\label{proof thm1}
	Based on the PDAs  obtained using {\textbf{Construction \ref{CON1}}}, a private coding scheme for a $\alpha-$connect PPMA-MADC model having  $KF$ mapper nodes and $K$ reducer nodes can be obtained as given below where we consider $Q={F \choose \alpha}$ output functions.
	\subsection{Map Phase}

	First, files are divided by grouping $N$ files into $\Lambda=KF$ disjoint batches $\{B_k^f: k\in [K],f\in [F]\}$, each containing $ \eta = \frac{N}{KF}$ files such that $\bigcup_{k \in [K], f\in [F]} B_{k}^f = \{W_1,W_2,\ldots,W_{N}\}$. The system consists of \(\Lambda=KF\) mapper nodes $\{M_k^f: k \in [K], f\in [F]\}$ organized into disjoint mapper node blocks of size $K$, $\M_k=\{M_k^f: f \in [F]\}$, for $k\in [K]$. There are  \(K \) reducer nodes $\{R_k: k\in [K]\}$, and \(Q ={F \choose \alpha}\) output functions $\{\phi_i, i \in [Q]\}$.  
	Each mapper node \(M_k^f\) stores exactly one batch \(B_k^f\) of files, for \(k \in [K], f\in [F]\).  
	Each reducer node is assigned a single output function to compute. Assume that reducer node \(R_k\) computes \(\phi_{d_k}\), for  \(k \in [K]\). The index $d_k$ is hidden from other reducer nodes, where $d_k \in [Q].$
	
	We introduce \((Q-1)K\) virtual reducer nodes alongside the \(K\) real reducer nodes, resulting in a total of \(KQ\) effective reducer nodes \(\{R_k^j : k\in [K], j \in [Q]\}\).  
	In the extended PDA \(\mathbf{P}\), each column corresponds to an effective reducer node, and each row corresponds to a batch of files (equivalently, a mapper node). A star \((\ast)\) in a given column indicates that the batch associated with that row is accessible at that effective reducer node.   
	The array \(\mathbf{P}\) is structured as a block matrix with \(K\) row blocks and \(K\) column blocks. Within each column block \(k \in [K]\), the \(j\)-th column corresponds to the effective reducer node \(R_k^j\), $j\in [Q]$.  Also, in each column block \(k\), exactly one column corresponds to the real reducer node \(R_k\); the remaining columns in that block represent virtual reducers.  
	Similarly, the row blocks of \(\mathbf{P}\) correspond to mapper node blocks: the first row block corresponds to mapper block \(\mathcal{M}_1\), the second row block to \(\mathcal{M}_2\), and so on.
	On the row side, row block \(k \in [K]\) represents the mapper node block \(\mathcal{M}_k\), and the \(f\)-th row inside that block represents mapper node \(M_k^f\), $f\in [F]$.  
Since \(M_k^f\) stores batch \(B_k^f\), each row also uniquely represents the file batch \(B_k^f\).

Each real reducer node \(R_k\) is connected to all mapper nodes in blocks \(\{\mathcal{M}_j : j \in [K] \setminus \{k\}\}\), but only to $\alpha$ mappers within its own block \(\mathcal{M}_k\).  
The $\alpha$-element subsets of \([F] = \{1,2,\ldots , F\}\) are ordered lexicographically.
Denote by \(y_{\alpha}(T)\) the index of the $\alpha-$subset \(T\) in this order.
Let the $\alpha$ mapper nodes in block \(\mathcal{M}_k\) to which \(R_k\) is connected be \(\{M_k^{t_i}: i \in [\alpha]\}\). Now, define
$a_k = y_{\alpha}(\{t_1,t_2,\dots, t_{\alpha}\}).$
The real reducer node \(R_k\) then corresponds to the \(a_k\)-th column within column block \(k\) of the extended PDA \(\mathbf{P}\).  
Inside \(\mathbf{P}\), considering the column block $k$, the subarray \({\bf P}^{(1)}\) (located in row block \(k\) and column block \(k\)) has stars exactly in rows \(t_1,t_2, \ldots, t_{\alpha}\) of column \(a_k\), which matches the connectivity of \(R_k\) to those mapper nodes and hence the accessibility of the corresponding file batches.

Thus, each real reducer \(R_k\) impersonates the effective reducer node \(R_k^{a_k}\).  
The value \(a_k\) is private to reducer \(R_k\) and unknown to the other reducers.
Each real reducer node \(R_k\) has access to file batches as follows:
$\mathcal{R}_k =\bigl\{B_f^j : f \in [K] \setminus \{k\},\; j \in [F] \bigr\} \cup  \bigl\{B_k^j : j \in T,\; T \subseteq [F],\; y_{\alpha}(T) = a_k,\; |T| = \alpha \bigr\}.$
Consequently, \(R_k\)  can access the following IVs: $
\{ v_{q,n} : q \in [Q],\; W_n \in B,\; B \in \mathcal{R}_k \}.$

	\subsection{Shuffle Phase}
		Each real reducer node \(R_k\) is responsible for computing the function \(\phi_k\).
	However, some of the required IVs are not locally available at \(R_k\).
	In particular, for real reducer node \(R_k\), the missing IVs are
	\begin{align}
	\{ v_{d_k,n} : W_n \in B, B \notin \R_k \}.
	\end{align}
	For each real reducer node $R_k,$ for $k\in [K]$ assigned to compute the output function indexed by  $d_k\in[Q]$ (i.e., the output function $\phi_{d_k}$), the following steps occur
	
	\begin{itemize}
		\item Real reducer node $R_k$ selects a vector $ {\bf y}_k  $ from all permutations of  $[Q]$ such that the $a_k$-th element is equal to $d_k$. The $i$-th entry in the vector \({\bf y}_k\)  represents  the output function index assigned to the effective reducer node $R_k^i$.
		\item Real reducer node $R_k$ broadcasts ${\bf y}_k$ to all the other real nodes. As a result, from the perspective of all other real reducer nodes, the union of the demands from the effective reducer nodes in  the set $ \{R_k^i, i \in [Q]\}$ always covers the entire set of output function indices $[Q]$, which is crucial for ensuring privacy.
	\end{itemize}  
	

	For each \(f\) such that ${\bf P}^{(1)}_{f,a_k} \neq *$, we aggregate the  IVs required by the effective reducer node \(R_k^{a_k}\) (which is impersonated by the real node \(R_k\)) that can be computed from the files in batch \(B_k^f\). This aggregated symbol is
	\[
	\U_{d_k, B_k^f} = \bigl(v_{d_k,n} : W_n \in B_k^f\bigr) \in \mathbb{F}_{2^{\eta \beta}},
	\]
	where \(\phi_{d_k} = \phi_{\mathbf{y}_k[a_k]}\) is the output function that \(R_k\) must compute.
	
	The symbol \(\U_{d_k, B_k^f}\) is then partitioned into \(K-1\) equally sized packets:
	\begin{align}
	\label{parti}
	\U_{d_k, B_k^f} = \bigl\{\U_{d_k, B_k^f}^{\,i} : i \in [K] \setminus \{k\}\bigr\}.
	\end{align}
	The integer entries in the PDA \(\mathbf{P}\) denotes the demands of the effective nodes. 
	More generally, for each \(j \in [Q]\) and each \(f\) such that ${\bf P}^{(1)}_{f,j} \neq *$, we form the aggregated symbol \(\U_{\mathbf{y}_k[j], B_k^f}\) containing all IVs that effective node \(R_k^j\) is assumed to compute from batch \(B_k^f\). This symbol is likewise split into \(K-1\) equal‑sized packets, each labeled with a distinct superscript from the set \([K] \setminus \{k\}\).
		\[
	\U_{\mathbf{y}_k[j], B_k^f} = \bigl\{\U_{\mathbf{y}_k[j], B_k^f}^{\,i} : i \in [K] \setminus \{k\}\bigr\}.
	\]
	The superscript assignment rule is that the demand corresponding to an integer entry in a given column block takes as superscripts the indices of the other five column blocks. 
	
	Let $\mathcal{J}_t^k = \{(j,f,i) : {\bf P}^{(1)}_{f,i} = t,\; j \in [K] \setminus \{k\},\; f \in [F], i \in [Q]\}.$
	Now, each real reducer node constructs coded transmissions as follows:  
	Real node \(R_1\) collects the packet parts of the demands that correspond to the integer \(1\) appearing in columns of all column blocks except block~1, XORs them, and broadcasts the result to the other real nodes.  
	The same procedure is repeated for every integer \(t \in [S]\) present in the PDA (since there are \(S ={F \choose \alpha +1}\) distinct integers in total).  
	Similarly, \(R_2\) processes the packet parts from all column blocks except block~2, and so on for every \(k \in [K]\).
	
	Formally, for each \(k \in [K]\) and for each \(t \in [S]\), real reducer node \(R_k\) produces the coded symbol
	\begin{align}
	X_k^t = \bigoplus_{(j,f,i) \in \mathcal{J}_t^k} \U_{\mathbf{y}_j[i], B_j^{f}}^{\,k},
	\end{align}
	and multicasts the sequence $\mathbf{X}_k = \{X_k^t : t \in [S]\}.$

	The real node $k$ can create the coded symbol $X_k^t$ from the IVs accessible to it. This follows since, for each column block indexed by $k_l \in [K] \backslash k$, the $2 \times 2$ sub-block array formed by row blocks $k, k_l$ and column blocks $k, k_l$ is either of the following forms
	$ \begin{pmatrix}
	{\bf P}^{(1)}& {\bf X}\\
	{\bf X} & {\bf P}^{(1)}
	\end{pmatrix} $or 
	$\begin{pmatrix}
	{\bf X}& 	{\bf P}^{(1)}\\
		{\bf P}^{(1)} & {\bf X}
	\end{pmatrix}.$ 
	\subsection{Reduce Phase}
	Receiving the sequences $\{{\bf X}_j\}_{j \in [K]\backslash k}$, each real node $R_k$ decodes all IVs of its output function, i.e., $\{v_{d_k,n} :  n \in [N])\}$
	with the help of IVs $\{v_{q,n} :  q \in [Q], W_n \in B, B\in \R_{k}\}$ it has access to, and finally computes the output function assigned to it.
	
	The real node $R_k$ needs to obtain $\{v_{d_k,n} :  W_n \in B, B\notin \R_{k}\}$, i.e., the set of IVs required for the output function $\phi_{d_k}$ from the files not accessible to it. Let $T_k$ represent the set of integers present in the $a_k-$th column within the column block $k$.
	Let ${\bf P}^{(1)}_{f,a_k} = t \in T_{k}$, for some $f\in [F]$. For each $k_l \in [K] \backslash k$ in (\ref{parti}), it can compute the symbol $\U_{d_k,B_k^{f}}^{k_l}$ from the coded symbol $X_{k_l}^t$ transmitted by the real node $k_l$, i.e., 
	\begin{align}
	\label{ transmission}
	X_{k_l}^t = \bigoplus_{(\hat{j},\hat{f},\hat{i})\in \J_t^{k_l}} \U_{{\bf y}_{\hat{j}}[\hat{i}],B_{\hat{j}}^{\hat{f}}}^{{k_l}}
	\end{align}
	for $\mathcal{J}_t^{k_l} = {(\hat{j},\hat{f},\hat{i}) : {\bf P}^{(1)}_{\hat{f},\hat{i}} = t,, \hat{j} \in [K] \setminus {k_l}, \hat{f} \in [F], \hat{i} \in [Q]}.$
	In (\ref{ transmission}), for $\hat{j} \neq k$, the $2 \times 2$ sub-block array formed by row blocks $k, \hat{j}$ and column blocks $k, \hat{j}$ is either of the following forms
	$ \begin{pmatrix}
	{\bf P}^{(1)}& {\bf X}\\
	{\bf X} & {\bf P}^{(1)}
	\end{pmatrix} $or 
	$\begin{pmatrix}
	{\bf X}&{\bf P}^{(1)}\\
	{\bf P}^{(1)} & {\bf X}
	\end{pmatrix}$. Hence, the real node $R_k$ can compute $\U_{{\bf y}_{\hat{j}}[\hat{i}],B_{\hat{j}}^{\hat{f}}}^{{k_l}}$ for each $\hat{j} \in [K] \backslash \{k, k_l\}$. If $\hat{j}=k$, then the real node can compute $\U_{{\bf y}_{\hat{j}}[\hat{i}],B_{\hat{j}}^{\hat{f}}}^{{k_l}}$ for each $\hat{i} \in [Q]\backslash a_k$ such that ${\bf P}^{(1)}_{\hat{f},\hat{i}}=t$, since ${\bf P}^{(1)}_{\hat{f},a_k}=*$ by the definition of PDA. For $\hat{i} = a_k$, ${\bf P}^{(1)}_{\hat{f},\hat{i}}={\bf P}^{(1)}_{{f},{a_k}} = t$ implies $\hat{f} = f$ by condition A3-1 of Definition \ref{def pda}. Therefore, the real reducer node $R_k$ can retrieve the symbol $\U_{d_k,B_k^f}^{k_l}$ from the coded symbol in (\ref{ transmission}) by canceling out the rest of the symbols (since $d_k= {\bf y}_k[a_k]$). By collecting all the symbols in (\ref{parti}), the real node $R_k$ can compute the output function $\phi_{d_k}$.
	
	Next, we compute the computation and communication loads for this scheme. Each real mapper node $\lambda \in [KF]$ stores $1$ batch of files. Hence, the computation load is
	$r=1$.

	Each reducer node send $S$ sequences  of size $\frac{\eta \beta}{(K-1)}$ bits by (\ref{ transmission}). 
	The communication load is given by
	\begin{align}
	L_p &= \frac{1}{QN\alpha} \frac{ K S\eta \alpha}{(K-1)} =  \frac{\eta }{\eta QKF } \frac{ KS }{(K-1)} 
	\nonumber\\&=  \frac{ S }{QF(K-1)} =  \frac{ {F \choose \alpha +1} }{{F \choose \alpha}F(K-1)} \nonumber \\
	&=  \frac{ {F - \alpha } }{{ (\alpha +1)}F(K-1)}. 
	\end{align}
	\subsection{Privacy}
	By our construction, using the PDAs, the allocation of file batches to each effective reducer node remains fixed. Consequently, $({\bf X}_1, \ldots, {\bf X}_{K})$ depend solely on the output functions assigned to the effective reducer nodes. From the perspective of other nodes, the value $a_j$, for $j \in [K]$, is uniform over $[Q]$.
	For any permutation ${\bf p}$ of $[Q]$, and for any $i \in [Q]$, $j,k \in [K]$ where $j \neq k$ (with $i$ being the $p$-th element of ${\bf p}$), the following holds
	\begin{align}
	& \Pr [ ({\bf y}_j[1], \ldots, {\bf y}_j[Q]) = {\bf p} \,|\, d_j = i, d_k, \mathcal{R}_k ] \nonumber \\
	&= \Pr [ ({\bf y}_j[1], \ldots, {\bf y}_j[Q]) = {\bf p} \,|\, d_j = i ] \label{eq:indep_other_info} \\
	&= \Pr [a_j = p \,|\, d_j = i ] \cdot \nonumber \\ & \quad \Pr [ ({\bf y}_j[1], \ldots, {\bf y}_j[p-1],  {\bf y}_j[p+1], \ldots, {\bf y}_j[Q]) \,|\, a_j \!=\! p, d_j \!=\! i ] \nonumber \\
	&= \frac{1}{Q} \cdot \nonumber \\ & \quad \Pr [({\bf y}_j[1], \ldots, {\bf y}_j[p-1],  {\bf y}_j[p+1], \ldots, {\bf y}_j[Q]) \,|\, a_j \!=\! p, d_j \!=\! i ] \label{eq:Sk_uniform} \\
	&= \frac{1}{Q} \cdot \frac{1}{(Q-1)!} \label{eq:uniformity_other_demand}
	= \frac{1}{Q!}.
	\end{align}
	where:
	\begin{itemize}
		\item \eqref{eq:indep_other_info} holds because, given $d_j$, the output function assignments of effective reducer nodes outside the range $\{R_j^m:m \in [Q]\}$ are independent of the stored batches and the output function assignments of other reducer nodes;
		\item \eqref{eq:Sk_uniform} holds because $a_j$ is uniformly distributed over $[Q]$, independently of $d_j$; and
		\item \eqref{eq:uniformity_other_demand} follows because, given $a_j$ and $d_j$, the output functions of effective reducer nodes in $\{R_j^m:m \in [Q]\}$ are uniformly selected from all permutations of $[Q]$, with the $a_j$-th element fixed as $d_j$.
	\end{itemize}
	For nodes outside the column block of reducer node $k$, i.e., reducer nodes not in $\{R_k^m:m \in [Q]\}$
	their permutations ${\bf y}_j$ depend only on their own $d_j$ and $a_j$, which are independent of $(d_k, \R_k)$. Thus, we have
	$H({\bf y}_j \mid d_k, \R_k) = H({\bf y}_j).$ From \eqref{eq:uniformity_other_demand}, we deduce that $\Pr [ {\bf y}_j \,|\, d_j, d_k, \mathcal{R}_k ]$ is independent of $(d_j, d_k, \mathcal{R}_k)$. This implies that the conditional mutual information can be expressed as follows:
	\begin{align}
	\label{indep}
	I({\bf y}_j ; d_j \mid d_k, \R_k) &= H({\bf y}_j \mid d_k, \R_k) - H({\bf y}_j \mid d_j, d_k, \R_k) \nonumber \\
	&= H({\bf y}_j ) - H({\bf y}_j ) = 0.
	\end{align}
	The mutual information can be decomposed as
	\begin{align}
	I({\bf y}_1, \ldots, {\bf y}_{K_1}; d_1, \ldots, d_{K_1} \mid d_k, \R_k) 
	= \nonumber \\   \sum_{j=1}^{K_1} I\big({\bf y}_j; d_j \mid d_k, \R_k\big),
	\end{align}
	because the output function assignments for nodes in different column blocks are independent given $d_k$ and $\R_k$.
	
	For any $j \ne k$, the output function assignments ${\bf y}_j$ for the effective nodes in column block $j$ are independent of $d_j$ given $d_k$ and $\R_k$ from (\ref{indep}).
	For $j = k$, since $d_k$ is already given, the mutual information is zero:
	\begin{align}
	I\big({\bf y}_k; d_k \mid d_k, \R_k\big) = 0.
	\end{align}
	Hence, we have
	\begin{align}
	I({\bf y}_1, \ldots, {\bf y}_{K}; d_1, \ldots, d_{K} \mid d_k, \R_k) 
	& = 0.
	\end{align}
	This shows that $\{{\bf y}_j\}_{j \in [K]}$ is independent of $\{d_j\}_{j \in [K]}$ when conditioned on $(d_k, \R_k)$. Therefore, the privacy constraint is satisfied, i.e., no node $k$ can infer any information about another node $j$'s output function index $d_j$ from the broadcast queries $\{{\bf y}_\ell\}_{\ell \in [K]}$, beyond what is already known from $(d_k, \R_k)$.
	
	\section{Proof of correctness of {\bf Algorithm 3}}
	\label{proof algo 3}
	We have seen in Section \ref{algo 3 overview} that {\bf Algorithm 3} constructs a $Q \times Q$ array with stars and integers in four systematic steps. Given integers $Q \geq 2$ and $\alpha < Q/2$, where both are either even or odd, the array is first initialized entirely with stars. In {\bf Step 2}, consecutive blocks of integers are placed row-wise in rows $\alpha+1$ to $m$, where $m = (Q+\alpha)/2$. {\bf Step 3} then creates symmetric repetitions by copying these integer rows to the lower half of the array (rows $m+1$ to $Q$) after applying systematically decreasing right cyclic shifts. Finally, in {\bf Step 4}, each column is independently cyclically shifted downward by an amount determined by its column index. The resulting array contains $Q(m-\alpha)$ distinct integers, each appearing exactly twice.
	

	
	We have seen that each integer in this array appears two times. Now we are going to prove Condition C3 in Definition 3.
	Let the first occurrence of $s$ (row-wise) be at position $(R, c)$ in the array before column shifts (before {\bf Step 4}), where $R$ is the original numeric row and $c$ is the column index. From {\bf Step 3}, we know that other occurrence of the integer $s$  is at row $R' = 2m - R + 1$ with column coordinate $c' \equiv c + (m - R) \pmod{Q}$, as determined by the right cyclic shift applied during 	\noindent {\textbf{Step 3}}. This establishes the fundamental relationship between the two occurrences before any column shifts are applied.

	After applying the column shifts in 	\noindent {\textbf{Step 4}}, each entry at position $(r, j)$ moves to position $((r + j - 1) \bmod Q, j)$. Therefore, the two occurrences of $s$ transform as follows.
	The first occurrence moves from $(R, c)$ to $(r_1, c)$ where $r_1 \equiv R + c - 1 \pmod{Q}$. The second occurrence moves from $(R', c')$ to $(r_2, c')$ where $r_2 \equiv R' + c' - 1 \pmod{Q}$. 
	Substituting $R' = 2m - R + 1$ and $c' \equiv c + m - R \pmod{Q}$ into the expression for $r_2$, we obtain 
	\[
	r_2 \equiv (2m - R + 1) + (c + m - R) - 1 \equiv 3m - 2R + c \pmod{Q}.
	\]
	Since $2m = Q + \alpha$, we have $3m = m + Q + \alpha$, which simplifies to
	\[
	r_2 \equiv m + \alpha - 2R + c \pmod{Q}.
	\]
	
	Now, we demonstrate that the other two positions in the $2 \times 2$ subarray formed by rows $\{r_1, r_2\}$ and columns $\{c, c'\}$, specifically positions $(r_1, c')$ and $(r_2, c)$, contain stars as entries in the final array.
	
	To determine the original location of an entry in the final array, we reverse the column shift transformation. For any position $(r, j)$ in the final array, its original row is given by $\text{orig\_row}(r, j) \equiv r - (j - 1) \pmod{Q}$.
	Now, for position $(r_1, c')$,
	\begin{align}
	\text{orig\_row}(r_1, c') \equiv r_1 - (c' - 1) &\equiv (R + c - 1) - (c' - 1) \nonumber \\&\equiv R + c - c' \pmod{Q}.
	\end{align}
	Substituting $c' \equiv c + m - R \pmod{Q}$ gives
	\[
	\text{orig\_row}(r_1, c') \equiv R - (m - R) \equiv 2R - m \pmod{Q}.
	\]
	For position $(r_2, c)$,
	\begin{align}
	\text{orig\_row}(r_2, c) \equiv r_2 - (c - 1) &\equiv (m + \alpha - 2R + c) - (c - 1) \nonumber \\ &\equiv m + \alpha - 2R + 1 \pmod{Q}.
	\end{align}
	
	We now analyze the ranges of $2R - m$ and $m + \alpha - 2R + 1$ modulo $Q$ for $R \in [\alpha +1, m]$.
	Since $R$ ranges from $\alpha +1$ to $m$, the expression $2R - m$ ranges from $2(\alpha +1) - m$ to $2m - m = m$. Given that $m = \frac{Q + \alpha}{2}$ and  both $Q$ and $\alpha$ are even or both are odd, we observe that $2R - m$ falls within the range $[1, \alpha]$ modulo $Q$ for all valid $R$.
	Similarly, $m + \alpha - 2R + 1$ ranges from $m + \alpha - 2m + 1 = \alpha - m + 1$ to $m + \alpha - 2(\alpha +1) + 1 = m - \alpha - 1$. Again, with $m = \frac{Q + \alpha}{2}$, this expression falls within $[1, \alpha]$ modulo $Q$.
	
	Since the original rows $1$ through $\alpha$ were initialized with stars and never assigned numeric values, both positions $(r_1, c')$ and $(r_2, c)$ originated from positions with entries as stars. The column shift operation in {\textbf{Step 4}} preserves the star positions (as it merely moves entries without changing their values), so these positions remain as stars in the final array.

	Therefore, the $2 \times 2$ subarray with rows $\{r_1, r_2\}$ and columns $\{c, c'\}$ contains $s$ at positions $(r_1, c)$ and $(r_2, c')$, and stars at positions $(r_1, c')$ and $(r_2, c)$. Depending on the relative ordering of $r_1$ and $r_2$, and $c$ and $c'$, this yields either the pattern $\begin{bmatrix} s & * \\ * & s \end{bmatrix}$ or $\begin{bmatrix} * & s \\ s & * \end{bmatrix}$.
	
	Hence all the conditions of PDA are satisfied by this array and hence, it is a $2-{\text{regular }}(Q,Q,\alpha, \frac{Q(Q-\alpha)}{2})$ PDA. 
	
	Note that from Section \ref{algo 3 overview}, before {\bf Step 4} (before column shifts), the first $\alpha$ rows are filled with stars and all other following rows are filled with integers. Now, column shifts and done in such a way that each column is shifted 1 units down compared to the previous column in a cyclic wrap around way, also in each column the stars appears in consecutive $\alpha$ positions. Hence, the PDA is $(1,2)-{\text{regular }}(Q,Q,\alpha, \frac{Q(Q-\alpha)}{2})$ PDA.

		\section{Proof of {\bf Construction 2}}
	\label{proof con 2}

For each $\alpha \in \left[1, \left\lceil \frac{Q}{2}\right \rceil \right ), $ where $Q+\alpha$ is  even, the PDA ${\bf P}^{(2)}$ in a $
(1,2)\text{-}\left ({Q}, {Q},\alpha,\frac{Q(Q-\alpha)}{2} \right )$ PDA.
	
	Each array \({\bf P} \in \mathcal{P}^{(2)}\) is structured as a block array consisting of \(K\) row blocks indexed by \(\left [{K} \right ]\) and \(K\) column blocks indexed by \([K]\). The block entries are denoted as \(\{{\bf p}_{f,k} : f, k \in [K]\}\). These blocks include the arrays \({\bf X}\) and \({\bf P}^{(2)} \), each of size \(Q \times Q \). Consequently, the overall size of \({\bf P}\) is \(KQ \times KQ \). 
	
	Each column in the PDA \({\bf P}^{(2)}\) contains exactly $\alpha$ stars, and  the array \({\bf X}\) contains exactly $Q$ stars per column. Thus, each column in \({\bf P}\) contains $\alpha +(K-1){Q }$ stars, satisfying condition \(A1\) of Definition \ref{def:PDA} with \(Z = \alpha +(K-1){Q }\). 
	The total number of integers in \({\bf P}\) is $\frac{Q(Q-\alpha)}{2} $. This satisfies condition \(A2\) of Definition \ref{def:PDA}.
	
	From \textbf{Construction \ref{CON1}}, for two distinct block entries \({\bf p}_{f_1,k_1}\) and \({\bf p}_{f_2,k_2}\), \({\bf p}_{f_1,k_1} = {\bf p}_{f_2,k_2} = {\bf P}^{(2)}\)  only if \(f_1 \neq f_2\) and \(k_1 \neq k_2\). These entries belong to distinct row and column blocks. Furthermore, \({\bf p}_{f_1,k_2} = {\bf p}_{f_2,k_1} = {\bf X}\), meaning the corresponding \(2 \times 2\) sub-block array formed by row blocks \(f_1, f_2\) and column blocks \(k_1, k_2\) must have one of the following forms:
	\[
	\begin{pmatrix}
	{\bf P}^{(2)} & {\bf X} \\
	{\bf X} & 	{\bf P}^{(2)} 
	\end{pmatrix}
	\quad \text{or} \quad
	\begin{pmatrix}
	{\bf X} & 	{\bf P}^{(2)}  \\
	{\bf P}^{(2)}  & {\bf X}
	\end{pmatrix}.
	\]

	In conclusion, the array \({\bf P}\) is a $\left ( KQ, KQ,(K-1)Q + \alpha,  \frac{Q(Q-\alpha)}{2}  \right ) $  PDA.
		\section{Proof of Theorem \ref{thm2}}
	\label{proof thm2}
	Based on the $(KQ,KQ,(K-1)Q+\alpha,\frac{Q(Q-\alpha)}{2})$ PDAs  obtained using {\textbf{Constriction \ref{CON2}}}, a private MADC scheme for a $\alpha-$cyclic PPMA-MADC model having  $KQ$ mapper nodes and $K$ reducer nodes can be obtained as given below where we consider $Q$ output functions.
	\subsection{Map Phase}
	
	First, files are divided by grouping $N$ files into $\Lambda=KQ$ disjoint batches $\{B_k^f: k\in [K],f\in [Q]\}$, each containing $ \eta = \frac{N}{KQ}$ files such that $\bigcup_{k \in [K], f\in [Q]} B_{k}^f = \{W_1,W_2,\ldots,W_{N}\}$. The system consists of \(\Lambda=KQ\) mapper nodes $\{M_k^f: k \in [K], f\in [Q]\}$ organized into disjoint mapper node blocks of size $K$, $\M_k=\{M_k^f: f \in [Q]\}$, for $k\in [K]$. There are  \(K \) reducer nodes $\{R_k: k\in [K]\}$, and \(Q \) output functions $\{\phi_i, i \in [Q]\}$.  
	Each mapper node \(M_k^f\) stores exactly one batch \(B_k^f\) of files, for \(k \in [K], f\in [Q]\).  
	Each reducer node is assigned a single output function to compute. Assume that reducer node \(R_k\) computes \(\phi_{d_k}\), for  \(k \in [K]\). The index $d_k$ is hidden from other reducer nodes.
	
	We introduce \((Q-1)K\) virtual reducer nodes alongside the \(K\) real reducer nodes, resulting in a total of \(KQ\) effective reducer nodes \(\{R_k^j : k\in [K], j \in [Q]\}\).  
	In the extended PDA \(\mathbf{P}\), each column corresponds to an effective reducer node, and each row corresponds to a batch of files (equivalently, a mapper node). An asterisk \((\ast)\) in a given column indicates that the batch associated with that row is accessible at that effective reducer node.   
	The array \(\mathbf{P}\) is structured as a block matrix with \(K\) row blocks and \(K\) column blocks. Within each column block \(k \in [K]\), the \(j\)-th column corresponds to the effective reducer node \(R_k^j\), $j\in [Q]$.  Also, in each column block \(k\), exactly one column corresponds to the real reducer node \(R_k\); the remaining columns in that block represent virtual reducers.  
	Similarly, the row blocks of \(\mathbf{P}\) correspond to mapper node blocks: the first row block corresponds to mapper block \(\mathcal{M}_1\), the second row block to \(\mathcal{M}_2\), and so on.
	On the row side, row block \(k \in [K]\) represents the mapper node block \(\mathcal{M}_k\), and the \(f\)-th row inside that block represents mapper node \(M_k^f\), $f\in [Q]$.  
	Since \(M_k^f\) stores batch \(B_k^f\), each row also uniquely represents the file batch \(B_k^f\).

	Each real reducer node \(R_k\) (where \(k \in [K]\)) impersonates one of the columns from column block \(k\) in the extended PDA, depending on the specific \(\alpha\) consecutive mapper nodes it is connected to within its own mapper node block \(\mathcal{M}_k\).  
If \(R_k\) is connected to mapper nodes in the circular interval $\{M_k^j: j \in [a_k, a_k+\alpha]_Q\}$ inside \(\mathcal{M}_k\), then it impersonates the \(a_k\)-th column of column block \(k\).  
Notice that in the PDA \(\mathbf{P}\), all entries of column block \(k\) in row blocks \(j \in [K] \setminus \{k\}\) are filled with the symbol \(\mathbf{X}\), indicating that reducer node \(R_k\) is connected to every mapper node in blocks \(\mathcal{M}_j\) for \(j \neq k\).  
Within row block \(k\) and column block \(k\), the subarray is the consecutive \(1\)-cyclic PDA ${\bf P}^{(2)}$.  

In \({\bf P}^{(2)}\), each column contains \(\alpha\) consecutive star entries: the first column has stars in rows $\{1,2,\ldots,\alpha\}$, the second column in rows $\{2,3,\ldots, \alpha +1\}$, and so on cyclically.  
Consequently, if \(R_k\) is connected to mapper nodes in the circular interval \([a_k, a_k+\alpha]_Q\) within \(\mathcal{M}_k\), it possesses access to batches \(\{B_k^{j}: j \in [a_k, a_k+\alpha]_Q\}\).  
This access pattern corresponds precisely to column \(a_k\) within column block \(k\) of the PDA, confirming that the PDA  models the problem.
	Thus, each real reducer \(R_k\) impersonates the effective reducer node \(R_k^{a_k}\).  
	The value \(a_k\) is private to reducer \(R_k\) and unknown to the other reducers.
	Each real reducer node \(R_k\) has access to file batches as follows:
	$\mathcal{R}_k =\bigl\{B_f^j : f \in [K] \setminus \{k\},\; j \in [Q] \bigr\} \cup  \bigl\{B_k^j :  j \in [a_k, a_k+\alpha]_Q\}.$
	Consequently, \(R_k\)  can access the following IVs: $
	\{ v_{q,n} : q \in [Q],\; W_n \in B,\; B \in \mathcal{R}_k \}.$
	
Rest of the proof follows from the Proof of Theorem \ref{thm1} in Section \ref{proof thm1}.

\end{document}